\documentclass[aps,prl,twocolumn,superscriptaddress,showpacs]{revtex4}
\usepackage{graphicx}
\usepackage{latexsym}
\usepackage{amssymb}
\usepackage{amsmath}
\usepackage{amsfonts}
\usepackage{bm}
\usepackage{multirow}
\usepackage{color}
\usepackage{comment}

\newcommand{\beq}{\begin{equation}}
\newcommand{\eeq}{\end{equation}}
\newcommand{\beqn}{\begin{eqnarray}}
\newcommand{\eeqn}{\end{eqnarray}}

\DeclareMathAlphabet{\mathbbold}{U}{bbold}{m}{n}

\makeatletter
\newcommand\xleftrightarrow[2][]{%
\ext@arrow 9999{\longleftrightarrowfill@}{#1}{#2}}
\newcommand\longleftrightarrowfill@{%
\arrowfill@\leftarrow\relbar\rightarrow} \makeatother

\begin{document}

\title{Coupled Wire description of the Correlated Physics in Twisted Bilayer Graphene}

\author{Xiao-Chuan Wu}
\affiliation{Department of Physics, University of California,
Santa Barbara, CA 93106, USA}

\author{Chao-Ming Jian}
\affiliation{Kavli Institute of Theoretical Physics, Santa
Barbara, CA 93106, USA} \affiliation{ Station Q, Microsoft
Research, Santa Barbara, California 93106-6105, USA}

\author{Cenke Xu}
\affiliation{Department of Physics, University of California,
Santa Barbara, CA 93106, USA}

\begin{abstract}

Since the discovery of superconductivity and correlated insulator
at fractional electron fillings in the twisted bilayer graphene,
most theoretical efforts have been focused on describing this
system in terms of an effective extended Hubbard model. However,
it was recognized that an exact tight binding model on the
Moir\'{e} superlattice which captures all the subtleties of the
bands can be exceedingly complicated. Here we pursue an
alternative framework of coupled wires to describe the system
based on the observation that the lattice relaxation effect is
strong at small twist angle, which substantially enlarges the AB
and BA stacking domains. Under an out-of-plane electric field
which can have multiple origins, the low energy physics of the
system is dominated by interconnected wires with (approximately)
gapless $1d$ conducting quantum valley hall domain wall states. We
demonstrate that the Coulomb interaction likely renders the wires
a $U(2)_2$ $(1+1)d$ conformal field theory with a tunable
Luttinger parameter for the charge $U(1)$ sector. Spin triplet and
singlet Cooper pair operator both have quasi-long range order in
this CFT. The junction between the wires at the AA stacking
islands can lead to either a two dimensional superconductor, or an
insulator.

\end{abstract}

\pacs{}

\maketitle


Surprising correlated physics such as superconductivity and
correlated insulator at fractional electron fillings away from
charge neutrality has been discovered in different systems with
Moir\'{e} superlattices~\cite{wangmoire,mag01,mag02,young2018},
which motivated a series of active theoretical
studies~\cite{xuleon,senthil,kivelson,fu,vafek,phillips,phillips2,baskaran,yang,louk,fu2,fu3,ashvinyou,martin,zhang,bernevig,scalet,senthil2,senthil3,fczhang,subir,balents2,dai}.
These systems have narrow electron bandwidth near charge
neutrality~\cite{flat1,flat2,flat3,flat4}, hence interaction
effects are significantly enhanced. In several systems that are
microscopically different, for example, (1) the heterostructure of
trilayer graphene (TLG) and hexagonal boron nitride (hBN), and (2)
twisted bilayer graphene (TBG), (3) twisted double bilayer
graphene (TDBG)~\cite{kimtalk}, insulating behavior was observed
at commensurate fractional fillings away from the charge neutral
point~\cite{wangmoire,mag01,young2018}; superconductivity has been
observed in all these systems near the insulator
phases~\cite{mag02,young2018,TLGSC,kimtalk}.

A consensus of the mechanism for the observed insulator and
superconductor has not yet been reached. A minimal triangular
lattice extended Hubbard model~\cite{xuleon} at least describes
the TLG/hBN heterostructure and twisted double bilayer graphene
with certain out-of-plane electric field (displacement
field)~\cite{senthil,band3,band4,band5}, since in these cases
there is no symmetry protected band touching below the fermi
energy, and the isolated narrow band has trivial topology. This
minimal model would then naturally predict either a
spin-triplet~\cite{xuleon} or spin-singlet $d+id$ topological
superconductor~\cite{kivelson}, depending on the sign of the
Hund's coupling. Signatures of spin triplet pairing predicted in
Ref.~\cite{xuleon} was recently found in TDBG~\cite{kimtalk},
though further experiments are demanded to determine the exact
pairing symmetry.

On the contrary, for one of the systems, $i.t.$ the TBG, it was
recognized that a standard tight binding model on the superlattice
that captures all the subtleties of the band structure can be
exceedingly complicated, and it may demand as many as ten bands
for each valley and each spin component~\cite{band1,band2}, which
makes analytical or numerical studies of this system very
difficult. These results suggest that an alternative theoretical
framework to understand the observed correlated physics is highly
desired for the TBG. Here we pursue a coupled wire network
framework to describe the TBG with a small twisted angle. A
similar description based on coupled wires, such as the
Chalker-Coddington model~\cite{CC,CC2}, has been used to describe
states without local Wannier orbitals. But in TBG, the coupled
wire network description is not just motivated by theoretical
convenience, it is also physically realistic, based on the
following observations:

(1) At small twisted angle, the lattice relaxation and deformation
effect is expected to be strong, and lead to substantially
enlarged AB and BA stacking domains~\cite{relax1,relax2}, and
narrow $1d$ domain walls.

(2) A displacement field will drive an AB (or BA) stacking bilayer
graphene into a ``quantum valley Hall
insulator"~\cite{vha,vhb,vhc,vhd,vhe,vhf,vhg}, and this
displacement field can be turned on manually
experimentally~\cite{young2018}, or intrinsically exists in the
system due to lack of $\hat{z} \rightarrow - \hat{z}$ reflection
symmetry (strongly asymmetric response to the displacement field
was indeed observed in Ref.~\cite{young2018}), or even be
generated spontaneously due to interaction~\cite{vh1}. Compared
with a single layer graphene, in an AB (or BA) stacking bilayer
graphene, interaction has much stronger effects due to the
quadratic band touching at each valley~\cite{vh2,vh3,vh4,vh5,qbt}.

(3) Under a uniform displacement field (regardless of its origin),
the AB and BA stacking domains are quantum valley Hall insulators
with opposite valley Hall conductivities, and they are separated
by domain walls with conducting $1d$ states. The long wavelength
modulation of the entire system prohibits large momentum transfer,
hence the valley quantum number is approximately conserved, and
the domain wall states are approximately gapless. These conducting
wires (AB/BA domain walls) have been observed directly in
numerics~\cite{networkn} and experiment on
TBG~\cite{relax3,network3}.

\begin{figure}
\includegraphics[width=200pt]{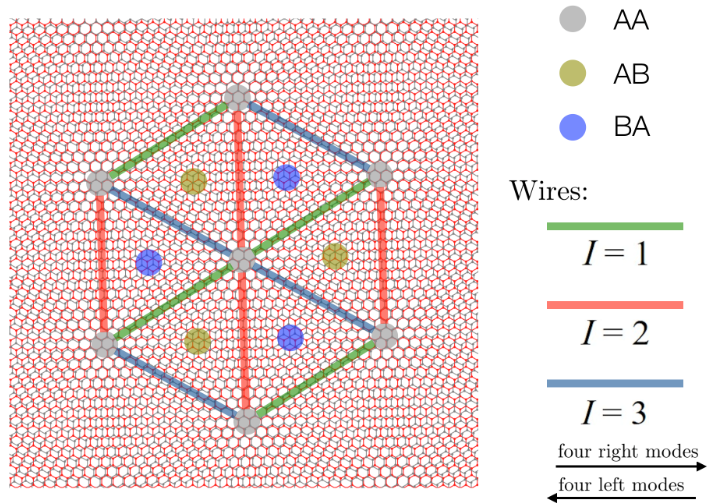}
\caption{The Moir\'{e} superlattice of TBG. If the lattice
relaxation and deformation effect is taken into
account~\cite{relax1,relax2,relax3}, the AB/BA stacking domains
would be substantially enlarged. There are four (two channels and
two spin components) left moving fermion modes and four right
moving modes along each wire (AB/BA domain wall). The left and
right moving fermions differ by a large lattice momentum
(orthogonal to the wires) which is the size of the Brillouin zone
of the original honeycomb lattice.} \label{fig}
\end{figure}

In fact, an effective network model has been proposed to describe
the noninteracting physics of the system~\cite{networkt}. In the
current work we will focus on the correlated phenomena. Along each
$1d$ wire, there are four counter-propagating localized electron
modes, which without interaction would constitute the $U(4)_1$
conformal field theory (CFT). The $1d$ fermions carry three
quantum numbers: valley ($L, R$), spin ($\uparrow, \downarrow$),
and channel ($1,2$) index (Fig.~\ref{fig}): \beqn H = \int dx \
\sum_{c = 1,2} \sum_{\alpha = \uparrow,\downarrow} i v
(\psi^\dagger_{L,c,\alpha}
\partial_x \psi_{L,c,\alpha} - \psi^\dagger_{R,c,\alpha}
\partial_x \psi_{R,c\alpha}). \eeqn
The left and right moving modes come from two different valleys
(which differ by a large momentum orthogonal to the wire), and
each valley will contribute two channels of chiral fermions, each
with two degenerate spin components. The displacement field in
experiment (for instance 0.5V/nm) corresponds to a much higher
energy scale compared with the sub kelvin environment of the
experiments. Thus we can safely assume that the quantum valley
Hall insulators are rather robust and these $1d$ wires, which form
a triangular lattice network, are dominating the low energy
physics.


The most important interaction in the system is still the Coulomb
interaction. 
The most noticeable effect of the Coulomb interaction is to
energetically favor two electrons to form a ``channel-singlet"
state, which is very similar to the mechanism of the standard
Hund's rule in transition metals. Let us consider two electrons
with the following two-body wave functions $\Psi_A(\mathbf{x}_1,
\mathbf{x}_2)$ and $\Psi_B(\mathbf{x}_1, \mathbf{x}_2)$
($\mathbf{x}_1$, $\mathbf{x}_2$ are $2d$ coordinates): \beqn
\Psi_A(\mathbf{x}_1, \mathbf{x}_2) &\sim&
\varphi_{L,1}(\mathbf{x}_1)\varphi_{R,2}(\mathbf{x}_2) -
\varphi_{L,2}(\mathbf{x}_1)\varphi_{R,1}(\mathbf{x}_2) \cr\cr &+&
\varphi_{R,1}(\mathbf{x}_1)\varphi_{L,2}(\mathbf{x}_2) -
\varphi_{R,2}(\mathbf{x}_1)\varphi_{L,1}(\mathbf{x}_2); \cr\cr
\Psi_B(\mathbf{x}_1, \mathbf{x}_2) &\sim&
\varphi_{L,1}(\mathbf{x}_1)\varphi_{R,2}(\mathbf{x}_2) -
\varphi_{L,2}(\mathbf{x}_1)\varphi_{R,1}(\mathbf{x}_2) \cr\cr &-&
\varphi_{R,1}(\mathbf{x}_1)\varphi_{L,2}(\mathbf{x}_2) +
\varphi_{R,2}(\mathbf{x}_1)\varphi_{L,1}(\mathbf{x}_2). \eeqn Here
$\varphi_{L,1}(\mathbf{x})$ represents the spatial wave function
of the left-moving fermions (which comes from one of the two
valleys) at channel 1. Both states $\Psi_{A,B}$ are ``channel
singlet" states (they are antisymmetric in the channel indices),
while $\Psi_A$ is symmetric in the valley space, $\Psi_B$ is
antisymmetric in the valley space. The spin space wave function
was not written down but can be straightforwardly inferred. Both
states cost low energy under Coulomb interaction, $i.e.$ they have
considerable lower energy compared with states that are symmetric
in the channel space, and this energy difference is not suppressed
by large momentum transfer (more detailed estimate will be given
in the supplementary material).
Thus the channel space is analogous to the gauged ``color space"
of spin chains~\cite{haldaneaffleck,affleck1986b}, which must form
a color singlet state.

A $U(4)_1$ CFT can be decomposed as \beqn U(4)_1 \sim U(1)^e_4
\oplus SU(2)^s_2 \oplus SU(2)^c_2, \eeqn where $SU(2)^c_2$
corresponds to the sector of the channel space. The interaction
effect discussed in the previous paragraph contributes to the
marginally relevant term $\lambda \mathcal{J}^c_L \cdot
\mathcal{J}^c_R$ in the CFT, where $\mathcal{J}^c_{L,R}$ are the
left and right Kac-Moody currents of the channel space, and it
will gap out the $SU(2)^c_2$ sector of the CFT. The residual
degrees of freedom would form CFT \beqn U(2)_2 \sim U(1)^e_4
\oplus SU(2)^s_2. \eeqn The $U(1)^e_4$ sector of the CFT
corresponds to the charge degrees of freedom, and it can be
represented by a pair of conjugate bosons $\theta$ and $\phi$
which satisfy $[\nabla_x \phi, \theta] = [\nabla_x \theta, \phi] =
i$. The $SU(2)^s_2$ corresponds to the spin degrees of freedom,
and as we discussed before, due to the prohibition of large
momentum transfer, the left and right modes have approximately
separate spin $SU(2)$ symmetries. The $SU(2)^s_2$ CFT can be
represented by a $(1+1)d$ nonlinear sigma model whose order
parameter is a $SU(2)$ matrix $g_{\alpha\beta}$, plus a
Wess-Zumino-Witten term at level-2~\cite{affleck1986}. The left
and right spin symmetry acts on $g_{\alpha\beta}$ as the left and
right $SU(2)$ transformations.

Physical operators can be represented as CFT fields. For example,
a fermion mass operator (which is a back-scattering term) can be
written as~\cite{affleck1986} \beqn \hat{M}_{\alpha\beta} = \sum_c
\psi^\dagger_{L,c,\alpha} \psi_{R,c,\beta} \sim \exp\left( i
\sqrt{\pi} \phi \right) g_{\alpha\beta}, \label{mass} \eeqn where
$g_{\alpha\beta}$ is the spin $SU(2)$ matrix order parameter
mentioned previously. Notice that the mass operator must be a
channel singlet, because otherwise it must involve the $SU(2)^c_2$
sector, which as we argued is already gapped out.

Likewise, a Cooper pair operator can be written as \beqn
\hat{\Delta}_{\alpha\beta} = \epsilon_{\alpha\gamma}\epsilon_{cd}
\psi_{L,c,\gamma} \psi_{R,d,\beta} \sim \exp\left( i \sqrt{\pi}
\theta \right) g_{\alpha\beta}, \label{cooper} \eeqn $\theta$ and
$\phi$ are the pair of conjugate bosons that describe the charge
sector of the CFT. The representation of the mass operator
$\hat{M}_{\alpha\beta}$ is given in Ref.~\cite{affleck1986}. The
Cooper pair operator representation can be inferred by defining a
new set of fermions: $\tilde{\psi}_L = \epsilon \epsilon
\psi_L^\dagger$, $\tilde{\psi}_R = \psi_R$, where the two
$\epsilon$ matrices act in the spin and channel indices
respectively. The fermion operator $\tilde{\psi}_{L}$ transforms
exactly the same as $\psi_L$ in the channel and spin space, but
carries opposite charge. The Cooper pair operator in
Eq.~\ref{cooper} becomes precisely the mass term (backscattering)
between $\tilde{\psi}_L$ and $\tilde{\psi}_R$.

The Cooper pair operator $\hat{\Delta}_{\alpha\beta}$ is a channel
singlet pairing. The pairing matrix $\hat{\Delta}_{\alpha\beta}$
can always be expanded as a four component vector $(\Delta^0,
\vec{\Delta})$: \beqn\hat{\Delta}_{\alpha\beta} = \Delta^0
\mathbf{1}_{2\times 2} + i \vec{\Delta} \cdot \vec{\sigma}. \eeqn
Here $\Delta^0$ is a spin singlet pairing order parameter, while
$\vec{\Delta}$ is a spin triplet pairing order parameter. Together
they form a four component vector representation under the $SO(4)
\sim SU(2)_L \times SU(2)_R$ symmetry. Without a further Hund's
(or anti-Hund's) coupling that favors either spin triplet or
singlet pairing, these four components pairing order parameters
are all degenerate. In the supplementary material, we discuss a
different method to obtain the CFT field expressions Eq.
\ref{mass} and Eq. \ref{cooper} where the fermion mass and the
Cooper pair operators are treated on equal footing.

The scaling dimensions of the fermion mass and Cooper pair
operators are \beqn [\hat{M}_{\alpha\beta}] = \frac{3}{8} +
\frac{1}{4K}, \ \ \ \ [\hat{\Delta}_{\alpha\beta}] = \frac{3}{8} +
\frac{K}{4}, \eeqn where $3/8$ comes from the scaling dimension of
the $g$ matrix order parameter in the $SU(2)^s_2$ CFT, and $K$ is
the Luttinger parameter in the $U(1)^e_4$ CFT. Soon we will see
that these scaling dimensions will determine whether the system
becomes superconductor or insulator due to wire junctions at the
AA islands. Notice that both $\hat{M}_{\alpha\beta}$ and
$\hat{\Delta}_{\alpha\beta}$ can simultaneously have lower scaling
dimensions (which implies enhanced correlation) compared with
noninteracting $1d$ fermion systems, where both operators have
scaling dimensions 1. Thus the interaction which gaps out the
$SU(2)^c$ channel space indeed enhances the system's tendency to
form superconductor and insulator.

The $U(1)^e_{4}$ CFT deserves some clarifications. It can always
be written as a free boson theory with the Hamiltonian: \beqn H =
\int dx \ \frac{1}{2K} (\nabla_x \theta)^2 + \frac{K}{2} (\nabla_x
\phi)^2. \label{luttinger}\eeqn $\theta$ and $\phi$ are a pair of
conjugate bosons. We can fermionize this theory through standard
procedure, and define new fermion operators as \beqn C_{L,R} \sim
\eta_{L,R} \exp( i \sqrt{\pi}\theta \pm i \sqrt{\pi} \phi),
\label{C} \eeqn where $\eta_{L,R}$ are the Klein factors. Then the
Cooper pair and the mass term of the new fermion $C_{L,R}$ should
be represented as $\exp(i \sqrt{4 \pi}\theta)$, and $\exp(i
\sqrt{4 \pi}\phi)$. But these Cooper pairs should correspond to
the charge$-4e$ bound state of the electrons, and the mass term
should correspond to a two electron backscattering. This is
because under the assumption of separate left and right spin
$SU(2)$ symmetries, a charge$-2e$ Cooper pair, or a singlet
electron back scattering term, cannot be invariant under the
$SU(2)_L \times SU(2)_R$ spin symmetry. Later we will show that
the charge$-4e$ $U(1)^e$ sector may become relevant to the finite
temperature physics of the system.

The $1d$ CFTs will intersect at the AA stacking islands, and due
to the lattice relaxation and deformation, the size of the AA
stacking islands has shrunk~\cite{relax1}. Let us first look at a
single AA island which is a junction between CFTs along three
directions. At this junction, the Cooper pairs can tunnel between
$1d$ CFTs along different wires. This Josephson tunnelling between
CFTs can be described by a $(0+1)d$ action at the junction \beqn
\mathcal{S} = \int d\tau \ \sum_{I,J} u_0 \Delta^{0
\dagger}_{\hat{e}_I} \Delta^0_{\hat{e}_J} + u_1
\vec{\Delta}^\dagger_{\hat{e}_I} \cdot \vec{\Delta}_{\hat{e}_J},
\label{S}\eeqn $\hat{e}_I$ with $I = 1,2,3$ represent wires along
three directions that meet at this junction. The scaling dimension
of $u_0$ and $u_1$ are both $ [u_0] = [u_1] = 1/4 - K/2$, where
$K$ is the Luttinger parameter in Eq.~\ref{luttinger}, thus when
$K < 1/2$ even a single junction Josephson Cooper pair tunnelling
becomes relevant, and we expect this Josephson tunnelling to drive
the entire system into a superconductor. If we take into account
of the tunnelling between parallel wires, which happens along the
entire $1d$ wires rather than one junction, then this parallel
tunnelling will be relevant and the entire system becomes a
superconductor for $K < 5/2$.

Here we allow $u_0$ and $u_1$ to be different, which breaks the
two separate $SU(2)$ spin symmetries to its diagonal spin $SU(2)$
symmetry. The AA island has shrunk substantially due to lattice
relaxation, thus the potential modulates at a shorter length scale
compared with other regions of the system, which enhances the
large momentum transfer and leads to the mixing between the left
and right $SU(2)$ symmetries. If $u_0$ dominate $u_1$, the system
would favor to form a global spin singlet pairing. Now the global
structure of the system can be mapped to the following classical
XY model: \beqn H &\sim& \sum_{\vec{r}} - V \sum_{I = 1}^3
\cos(\theta^I_{\vec{r}} - \theta^I_{\vec{r} + a \hat{e}_I}) \cr\cr
&+& u_0 \sum_{I,J=1}^3 \cos(\theta^I_{\vec{r}} -
\theta^J_{\vec{r}}) + \cdots \label{H} \eeqn Here $\vec{r}$ denote
the AA stacking islands of the lattice, and $\hat{e}_I$ with $I =
1,2,3$ are unit vectors along the wires (Fig.~\ref{fig}). $a$ is
the distance between two AA stacking islands, and
$\theta^I_{\vec{r}}$ is the phase angle of the spin singlet Cooper
pair of wire along direction $\hat{e}_I$. The ellipsis in
Eq.~\ref{H} represent other weaker terms allowed by symmetry in
the system

Here naturally $V > 0$, which reflects the fact that along each
wire the superconductor order parameter has a quasi long range
order and prefers the Cooper pair to have a uniform pairing phase
along the wire. Then when $u_0 < 0$, the Josephson couplings
between different wires are ``unfrustrated", hence the entire
system should form a spin singlet $s-$wave pairing with a uniform
pairing phase; while when $u_0 > 0$, the Josephson coupling
between wires along three directions is ``frustrated". The two
terms in Eq.~\ref{H} demands a uniform $\theta^I$ along direction
$\hat{e}_I$, while wires that intersect each other at one island
will have Cooper pair phases which differ from each other by $\pm
120$ degrees. Then the pairing symmetry of the entire system is
identical to the $d+id$ (or $d - id$) pairing, as under a spatial
60 degree rotation (a cyclic permutation between wires along three
directions), the pairing phase angle changes by $\pm 120$ degrees.
This $d+id$ pairing superconductor is a singlet of spin, valley,
and channel indices.

\begin{table}
\begin{tabular}{|c|c|c|}
\hline wires  & $u_0 < 0$, $s$-wave pairing & $u_0 > 0$, $d + id$
or $d-id$ pairing\tabularnewline \hline $I=1$ & $\Delta$ &
$\Delta$\tabularnewline \hline $I=2$ & $\Delta$ & $\Delta e^{\pm
i\frac{2\pi}{3}}$\tabularnewline \hline $I=3$ & $\Delta$ & $\Delta
e^{\mp i\frac{2\pi}{3}}$\tabularnewline \hline
\end{tabular}
\caption{The SC order parameter along different wires, with $u_0 <
0$ and $u_0 > 0$ in Eq.~\ref{H}. The index $I$ refers to the wires
in Fig.~\ref{fig}.}\label{table}
\end{table}

When $u_1$ dominates $u_0$ in Eq.~\ref{S}, the system will form a
spin triplet superconductor. As an example let us assume that
$\vec{\Delta}_{\hat{e}_I}(\vec{r}) = \exp(i \theta^{I}_{\vec{r}})
\vec{n}^I_{\vec{r}}$ (the real and imaginary parts of the spin
triplet Cooper pair are parallel with each other), which is
similar to the so called ``polar state" of Bose-Einstein
condensate (BEC) of the spin-1 spinor cold
atoms~\cite{polar1,polar1b,polar2}. Then the effective Hamiltonian
of the coupled Josephson wires reads \beqn H \sim \sum_{\vec{r}}
&-& V \sum_{I = 1}^3 \vec{n}^I_{\vec{r}} \cdot \vec{n}^I_{\vec{r}
+ a \hat{e}_I} \cos(\theta^I_{\vec{r}} - \theta^I_{\vec{r} + a
\hat{e}_I}) \cr\cr &+& u_1 \sum_{I,J=1}^3 \vec{n}^I_{\vec{r}}
\cdot \vec{n}^J_{\vec{r}} \cos(\theta^I_{\vec{r}} -
\theta^J_{\vec{r}}) + \cdots \label{H2} \eeqn When $u_1 < 0$, the
system forms a uniform $s-$wave spin triplet pairing. When $u_1 >
0$, again the Josephson coupling on every AA island is frustrated,
then the system either forms a uniform state of $\theta$, with a
120 degree ``antiferromagnetic" pattern of $\vec{n}$, or forms a
$d+id$ pattern of $\theta$, with a ``ferromagnetic" state of
$\vec{n}$. Other symmetry allowed terms, or quantum fluctuation
may lift the degeneracy of the two scenarios described above.

There is a $Z_2$ gauge transformation shared between $\exp(i
\theta^{I}_{\vec{r}})$ and $ \vec{n}^I_{\vec{r}}$, $i.e.$ the spin
triplet pairing order parameter is invariant under $
\vec{n}^I_{\vec{r}} \rightarrow - \vec{n}^I_{\vec{r}}$ and
$\theta^{I}_{\vec{r}} \rightarrow \theta^{I}_{\vec{r}} + \pi$. At
any finite temperature, the vectors $\vec{n}^I_{\vec{r}}$ will be
disordered due to thermal fluctuation because this system is
purely two dimensional, then as was predicted in
Ref.~\cite{xuleon}, the superconductor vortex at finite
temperature will carry magnetic flux quantized as $n hc/(4e)$.
This means that the charge sector will form an effective
charge$-4e$ superconductor with algebraic correlation of
charge$-4e$ order parameters. This charge-$4e$ superconductor is
qualitatively the same as the Cooper pair of the fermions
$C_{L,R}$ defined before. The same logic led to fractionalized
vortices of the polar state of spin-1 BEC, which was confirmed
numerically in Ref.~\cite{polar2}.



At the AA islands, symmetry also allows charge backscattering
within each wire. The charge sector of the system is described by
the $C$ fermions defined in Eq.~\ref{C}. $C_L$ and $C_R$ come from
two different valleys in the bulk, which project to the same
momentum (Dirac crossing) along the $1d$ domain wall. Upon doping
away from charge neutrality, the $C_{L,R}$ fermion will acquire a
fermi wave vector $\pm \delta k_f$ away from the Dirac crossing,
thus a backscattering involves a momentum transfer of $2 \delta
k_f$. The backscattering of the $C$ fermion is described by \beqn
\mathcal{S} = \int d\tau dx \ u U(x) \left( C^\dagger_L C_R e^{2i
\delta k_f x} + H.c. \right) \eeqn where $U(x)$ is the periodic
potential along the wire due to the AA stacking islands. If the
integral along the entire wire $\int dx U(x) e^{i2 \delta k_f x} $
is nonzero, then this implies that $2 \delta k_f = \pm 2\pi/a$,
where $a$ is the lattice constant of the Moir\'{e} superlattice,
or the distance between two AA stacking islands. This implies that
there must be extra integer multiple of $\pm 2e$ charges between
two AA islands on each wire (one $C$ fermion carries charge $2e$).
And if wires along two directions acquire $+2e$ between every two
neighboring AA islands, and the wires along the third direction
acquire $-2e$ between AA islands, the entire system becomes an
insulator at half-filling away from charge neutrality with $+2e$
charge per unit cell on the superlattice. The insulator observed
at the $1/4$ filling should correspond to two particle
backscattering, which is a much weaker effect. The backscattering
will be more relevant with larger Luttinger parameter $K$.


We also notice that in experiment the resistivity at the same
charge density can strongly depend on the displacement
field~\cite{young2018}. This is a natural phenomenon in our
formalism, because a stronger displacement field would lead to a
larger gap in the quantum valley Hall insulator, and hence
stronger localization of the electron wave function at the wires.
Stronger localization of the domain wall states would lead to a
stronger effective particle density-density interaction in the
$(1+1)d$ CFT, and hence a larger Luttinger parameter $K$ based on
the standard bosonization formalism. A larger $K$ would render the
backscattering at the AA islands more relevant. This means that
the Luttinger parameter $K$ is tunable by the displacement field,
and the field can potentially lead to a metal-insulator
transition.

{\it Summary:} We study the correlated physics of the TBG based on
a coupled wire framework. The low energy physics of the system is
dominated by the conducting wires which are the domain walls
between the AB/BA domains. These domains are enlarged due to
lattice relaxation, and are driven into the quantum valley Hall
insulators under a displacement field which can have multiple
origins. The observed superconductivity and the correlated
insulator of the system are interpreted as consequences of the
Josephson tunnelling and also backscattering at the AA stacking
islands, which are the junctions where the wires along three
directions meet. One puzzle from the experiment is the weakness of
the insulators at fractional fillings. In our description, the
insulating behavior is due to the backscattering at the AA
islands, which is still suppressed due to large momentum transfer
(large momentum transfer orthogonal to the wire, which is still
approximately defined due to the smoothness of the background
potential), thus it will at most lead to a weak correlated
insulator. In our formalism a displacement field can tune the
Luttinger parameter of the CFT, and hence affect the relevance of
backscattering and also charge transport, as was observed
experimentally.

Chao-Ming Jian's research at KITP is supported by the Gordon and
Betty Moore Foundations EPiQS Initiative through Grant GBMF4304.
Cenke Xu is supported by the David and Lucile Packard Foundation.

\bibliography{network}

\appendix

\section{A: Exchange energy of two-particle wave functions}

Let us evaluate the exchange energy of two-particle wave functions
in more detail in this appendix. The wave function
$\Psi_A(\mathbf{x}_1, \mathbf{x}_2)$ considered in the main tex
has the interaction energy \beqn E_{int} &\sim& \int d\mathbf{x}_1
d\mathbf{x}_2 \ \Psi^\ast_A(\mathbf{x}_1, \mathbf{x}_2)
V_{\mathbf{x}_1,\mathbf{x}_2} \Psi_A(\mathbf{x}_1, \mathbf{x}_2)
\cr\cr &=& E_0 + E_{ex}; \eeqn where
$V_{\mathbf{x}_1,\mathbf{x}_2}$ is the (screened) Coulomb
interaction. Both integrals $\int d\mathbf{x}_1$, $\int
d\mathbf{x}_2$ are performed in the $2d$ space. \beqn E_0 = \int
d\mathbf{x}_1 d\mathbf{x}_2 |\varphi_{L,1}(\mathbf{x}_1)|^2
|\varphi_{R,2}(\mathbf{x}_2)|^2 V_{\mathbf{x}_1,\mathbf{x}_2} +
\cdots \eeqn $E_{ex}$ is the exchange energy, and it involves six
integrals: \beqn
 && I_{ex,1} \sim -
 \cr && \int d\mathbf{x}_1 d\mathbf{x}_2 \varphi^\ast_{L,1}(\mathbf{x}_1)\varphi_{L,2}(\mathbf{x}_1) V_{\mathbf{x}_1,\mathbf{x}_2} \varphi^\ast_{R,2}(\mathbf{x}_2)\varphi_{R,1}(\mathbf{x}_2) + c.c;
 \cr\cr
 && I_{ex,2} \sim +
 \cr && \int d\mathbf{x}_1 d\mathbf{x}_2 \varphi^\ast_{L,1}(\mathbf{x}_1)\varphi_{R,1}(\mathbf{x}_1) V_{\mathbf{x}_1,\mathbf{x}_2} \varphi^\ast_{R,2}(\mathbf{x}_2)\varphi_{L,2}(\mathbf{x}_2) + c.c;
 \cr\cr
 && I_{ex,3} \sim -
 \cr && \int d\mathbf{x}_1 d\mathbf{x}_2 \varphi^\ast_{L,1}(\mathbf{x}_1)\varphi_{R,2}(\mathbf{x}_1) V_{\mathbf{x}_1,\mathbf{x}_2} \varphi^\ast_{R,2}(\mathbf{x}_2)\varphi_{L,1}(\mathbf{x}_2) + c.c;
 \cr\cr
 && I_{ex,4} \sim -
 \cr && \int d\mathbf{x}_1 d\mathbf{x}_2 \varphi^\ast_{L,2}(\mathbf{x}_1)\varphi_{R,1}(\mathbf{x}_1) V_{\mathbf{x}_1,\mathbf{x}_2} \varphi^\ast_{R,1}(\mathbf{x}_2)\varphi_{L,2}(\mathbf{x}_2) + c.c;
 \cr\cr
 && I_{ex,5} \sim +
 \cr && \int d\mathbf{x}_1 d\mathbf{x}_2 \varphi^\ast_{L,2}(\mathbf{x}_1)\varphi_{R,2}(\mathbf{x}_1) V_{\mathbf{x}_1,\mathbf{x}_2} \varphi^\ast_{R,1}(\mathbf{x}_2)\varphi_{L,1}(\mathbf{x}_2) + c.c;
 \cr\cr
 && I_{ex,6} \sim -
 \cr && \int d\mathbf{x}_1 d\mathbf{x}_2 \varphi^\ast_{R,1}(\mathbf{x}_1)\varphi_{R,2}(\mathbf{x}_1) V_{\mathbf{x}_1,\mathbf{x}_2} \varphi^\ast_{L,2}(\mathbf{x}_2)\varphi_{L,1}(\mathbf{x}_2 + c.c;
 \cr\cr
 && E_{ex} = \sum_{i = 1}^6 I_{ex, i}, \eeqn
The single-particle wave functions are roughly (for example)
$\varphi_{L,1}(\mathbf{x}) \sim \exp(i K_{1} x) F_{L,1}(y)$, etc.
where $F_{L,1}(y)$ is an envelop function of the coordinate $y$
orthogonal to the wire, and localized at the wire. $F_{L,1}(y)$
should carry an approximately conserved large momentum, which
inherits from the crystal momentum of one of the two valleys,
assuming the domain wall is smooth enough compared with the
lattice scale. In all these exchange energy integrals, $I_{ex,2 -
5}$ are expected to be considerably smaller than $I_{ex,1}$ and
$I_{ex,6}$, because they involve large momentum transfer, $i.e.$
integrals like $\int d\mathbf{x}_1
\varphi^\ast_{L,1}(\mathbf{x}_1)\varphi_{R,1}(\mathbf{x}_1)$,
These integrals are highly suppressed because
$\varphi_{L,1}(\mathbf{x}_1)$ and $\varphi_{R,1}(\mathbf{x}_1)$
come from two valleys in the original honeycomb lattice, the two
valleys have very large momentum difference.

\begin{figure}
\includegraphics[width=190pt]{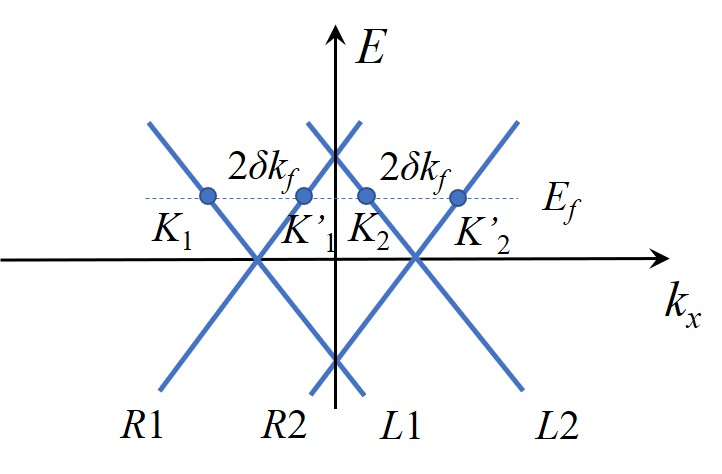}
\caption{Schematic dispersion of the $1d$ domain wall states after
doping. $K_{1}$ and $K_{2}$ come from the same valley $\mathbf{Q}$
in the $2d$ Brillouin zone. Time-reversal symmetry guarantees that
$K_{1} = - K'_{2}$, $K_{2} = - K'_{1}$. } \label{1d}
\end{figure}

$I_{ex,1} + I_{ex,6}$ is the main exchange energy gained by
$\Psi_A$, both integrals do not involve large momentum transfer,
and they both conserve the total momentum along the wire
(time-reversal symmetry guarantees that $K_{1} = - K'_{2}$, $K_{2}
= - K'_{1}$), assuming we focus on a single wire without junction.
With the Coulomb interaction, or the standard form of screened
Coulomb interaction, $I_{ex,1} + I_{ex,6}$ is negative. The
exchange energy of $\Psi_B$ is very similar, and both wave
functions are ``channel" singlet states.

One can also run the same test on other two-particle wave
functions which are symmetric in the channel space, such as \beqn
\Psi_C(\mathbf{x}_1, \mathbf{x}_2) &\sim&
\varphi_{L,1}(\mathbf{x}_1)\varphi_{R,2}(\mathbf{x}_2) +
\varphi_{L,2}(\mathbf{x}_1)\varphi_{R,1}(\mathbf{x}_2) \cr\cr &-&
\varphi_{R,1}(\mathbf{x}_1)\varphi_{L,2}(\mathbf{x}_2) -
\varphi_{R,2}(\mathbf{x}_1)\varphi_{L,1}(\mathbf{x}_2); \cr\cr
\Psi_D(\mathbf{x}_1, \mathbf{x}_2) &\sim&
\varphi_{L,1}(\mathbf{x}_1)\varphi_{R,1}(\mathbf{x}_2) -
\varphi_{L,2}(\mathbf{x}_1)\varphi_{R,2}(\mathbf{x}_2) \cr\cr &+&
\varphi_{R,1}(\mathbf{x}_1)\varphi_{L,1}(\mathbf{x}_2) -
\varphi_{R,2}(\mathbf{x}_1)\varphi_{L,2}(\mathbf{x}_2); \cr\cr
\cdots\cdot \eeqn None of these wave functions gain as much
exchange energy compared with $\Psi_A$ and $\Psi_B$, because their
exchange energy integrals either involve large momentum transfer,
or violate total momentum conservation along the wire. For
example, for $\Psi_D(\mathbf{x}_1, \mathbf{x}_2)$, its exchange
energy contains terms like \beqn - \int d\mathbf{x}_1
d\mathbf{x}_2 \varphi^\ast_{L,1}(\mathbf{x}_1)
\varphi_{L,2}(\mathbf{x}_1) V_{\mathbf{x}_1,\mathbf{x}_2}
\varphi^\ast_{R,1}(\mathbf{x}_2) \varphi_{R,2}(\mathbf{x}_2),
\eeqn this integral represents the physical process of moving two
particles at momenta $K_{2}$ and $K'_{2}$ to momenta $K_{1}$ and
$K'_{1}$ (Fig.~\ref{1d}), which is suppressed because in general
it violates total momentum conservation along the wire. This total
momentum conservation can be viewed as a $U(1)$ symmetry in the
channel space, $i.e.$ $N_{L,1} + N_{R,1} - N_{L,2} - N_{R,2}$ must
be a conserved quantity, where (for example) $N_{L,1}$ is the
number of left moving particles at channel 1.

\section{B: Fermion Bilinears as CFT fields}

In the main text, we obtain the CFT field expressions of the
fermion mass operator Eq.~5 using the non-Abelian bosonization of
$U(4)_1$ and the decomposition $U(4)_1 \sim U(1)_4 \otimes
SU(2)_2^s \otimes SU(2)_2^c$. For the Cooper pair operator Eq.~6,
we first define a new basis of fermions such that the Cooper pair
operator acts as a fermion mass operator in the new basis. Then,
we conduct a similar non-Abelian bosonization and the
decomposition of $U(4)_1$ to obtain its CFT field expression. In
this section, we study a different method to obtain the CFT field
expressions of the fermion mass operator and the Cooper pair
operator while treating them in equal footing.

We first rewrite the left/right-moving complex fermions
$\psi_{L,c,\alpha}$, $\psi^\dag_{L,c,\alpha}$,
$\psi_{R,c,\alpha}$, and $\psi^\dag_{R,c,\alpha}$ in a Majorana
fermion basis $\chi_{L/R}$ where each of $\chi_L$ and $\chi_R$ is
an 8-component Majorana spinor. The Majorana fermion basis is
chosen such that the generators of the symmetries $U(1)^e$,
$SU(2)^s$ and $SU(2)^c$ are given by \beqn && U(1)^e:
\sigma^{y00},
\nonumber\\
&& SU(2)^s: \sigma^{0xy},~\sigma^{0zy},~\sigma^{0y0},
\label{SU(2)^c generators}
\\
&& SU(2)^c: \sigma^{0yx},~\sigma^{0yz},~\sigma^{00y}, \nonumber
\eeqn where $\sigma^{x,y,z}$ are the Pauli matrices and $\sigma^0$
is the $2\times 2$ identity matrix. Here, we've adopted the
notation $\sigma^{abc...}\equiv \sigma^a \otimes \sigma^b \otimes
\sigma^c \otimes ...$. The left and right-moving Majorana fermions
can be described by the $O(8)_1$ CFT. More precisely, we can
bosonize these Majorana femions and describe them using a
non-linear sigma model with the group $O(8)$ and with a
Wess-Zumino-Witten term at level 1. Following the non-Abelian
bosonization procedure given by Ref.~\onlinecite{witten1984}, we
can identify the fermion bilinears $\chi_L \chi_R^T$ with the
field $h\in O(8)$ of the non-linear sigma model. The fermion mass
operator in Eq.~5 and the Cooper pair operator Eq.~6 are included
in $\chi_L \chi_R^T$ and hence can be expressed in terms of $h\in
O(8)$ when bosonized. In the following, we will study the specific
form of field $h\in O(8)$ which represents the fermion mass and
the Cooper pair operators.

First of all, both of the fermion mass and the Cooper pair
operators are $SU(2)^c$ singlets. Hence, we focus only on the
field $h\in O(8)$ such that $h$ commutes with the $SU(2)^c$
generators given in Eq. \ref{SU(2)^c generators}. The field $h$
that satisfy this condition takes the general form
\begin{align}
h = W (\tilde{h} \otimes \sigma^0) W^\dag \label{h_field}
\end{align}
where $W= \frac{1}{\sqrt{2}} ( 1 + i \sigma^{0yy})$ and
$\tilde{h}$ is a $4\times 4$ matrix. Since $h \in O(8)$, $h$ has
to be a real matrix, which implies \beqn \sigma^{0y} \tilde{h}
\sigma^{0y} = \tilde{h}^*. \eeqn This condition implies that
$\tilde{h}$ decompose into a linear superposition of the following
basis matrices with real coefficients: \beqn && \sigma^{00},~
i\sigma^{0x}, ~i\sigma^{0y},~ i\sigma^{0z}
\nonumber \\
&& i\sigma^{y0},~ \sigma^{yx}, ~\sigma^{yy},~ \sigma^{yz}
\nonumber \\
&& \sigma^{x0},~ i\sigma^{xx}, ~i\sigma^{xy},~ i\sigma^{xz}
\nonumber \\
&& \sigma^{z0},~ i\sigma^{zx}, ~i\sigma^{zy},~ i\sigma^{zz}.
\label{Eq:basis_h} \eeqn Both the fermion mass and the Cooper pair
operators transform non-trivially under the left and right
$U(1)^e$ and $SU(2)^s$ actions. For the field $h$, the left and
right $U(1)^e$ and $SU(2)^s$ actions are given by the left and
right multiplication of $U(1)^e$ and $SU(2)^s$ matrices generated
the generators given in Eq. \ref{SU(2)^c generators}. Hence, we
should organize the basis of $\tilde{h}$ such that $h$ transforms
properly under the left and right $U(1)^e$ and $SU(2)^s$ actions:
\beqn \tilde{h} &=& \alpha (\cos \phi + i \sin\phi \sigma^y)
\otimes g \cr\cr &+& \beta (\cos \theta \sigma^x +  \sin\theta
\sigma^z) \otimes g', \eeqn where $\alpha,\beta$ are real number,
$\phi$ and $\theta$ are two angular variables, and $g,g'\in SU(2)$
are $2\times 2$ $SU(2)$ matrices. Note that $\tilde{h}$ contains
two terms. Their transformations under the left and right $U(1)^e$
symmetries allow us to identify them as the fermion mass operator
and the Cooper pair operators respectively. The angular variables
$\phi$ and $\theta$ are then naturally identify with the $\phi$
and $\theta$ fields of the $U(1)_4^e$ CFT fields discussed in the
main text. Finally, we need to consider the constrain of $h^T h =
1$ on $\tilde{h}$: \beqn
&& \alpha^2 +\beta^2 =1, \nonumber \\
&& g g' = g' g. \eeqn To treat the fermion mass operator and the
Cooper pair operator in equal footing, we should choose
$\alpha=\beta=\frac{1}{\sqrt{2}}$. The second equation is
naturally satisfy by setting $g=g' \in SU(2)$. Now, we can
conclude that the most generic form of $\tilde{h}$ that captures
the fermion mass operators and Cooper pair operators in equal
footing is given by \beqn \tilde{h} &=& \frac{1}{\sqrt{2}}(\cos
\phi + i \sin\phi \sigma^y) \otimes g \cr\cr &+&
\frac{1}{\sqrt{2}} (\cos \theta \sigma^x +  \sin\theta \sigma^z)
\otimes g. \eeqn Using this form of $\tilde{h}$, we can obtain the
expression of $h$. We can furthermore transform the basis from
$\chi_{L/R}$ back to the complex fermions $\psi_{L,c,\alpha}$,
$\psi^\dag_{L,c,\alpha}$, $\psi_{R,c,\alpha}$. After the basis
transformation, we see that the two terms in $h$ (that comes from
the two terms in $\tilde{h}$) agree respectively with the CFT
field expressions of the fermion mass operator Eq.~5 and of the
Cooper pair operator Eq.~6 in the main text.

\end{document}